\documentclass[letterpaper, 10 pt, conference]{ieeeconf}  % Comment this line out if you need a4paper

\IEEEoverridecommandlockouts                              % This command is only needed if 
\usepackage{multirow}

\usepackage{amsmath}

\usepackage{algorithmic}
\usepackage{graphicx}
\usepackage{textcomp}

\usepackage{tabularx,colortbl}

\usepackage{bbold}

\usepackage[flushleft]{threeparttable}

 \usepackage{hyperref}
 \hypersetup{
     colorlinks=true,
     linkcolor=blue,
     filecolor=blue,
     citecolor=green,
     urlcolor=cyan,
}

\title{\LARGE \bf
Transformer Convolutional Neural Networks for Automated Artifact Detection in Scalp EEG
}

\author{Wei Yan Peh$^{1,2*}$, Yuanyuan Yao$^{3}$, and Justin Dauwels$^{3}\dagger$% <-this % stops a space
\thanks{This is an extension to a paper presented at the 2022 44th Annual International Conference of the IEEE Engineering in Medicine \& Biology Society (EMBC) Scottish Event Campus, Glasgow, UK, July 11-15, 2022.}
\thanks{$^{1}$ Nanyang Technological University (NTU), Interdisciplinary Graduate Programme (IGP), Singapore 639798.}
\thanks{$^{2}$ NTU, School of Computer Science and Engineering (SCSE), Singapore 639798.}
\thanks{$^{3}$ Department of Microelectronics, Delft University of Technology, 2628 CD Delft, Netherlands}
\thanks{* corresponding author email: pehw0012@e.ntu.edu.sg}
\thanks{$\dagger$ corresponding author email: j.h.g.dauwels@tudelft.nl}
}

\begin{document}

\maketitle
\pagestyle{empty}

%%%%%%%%%%%%%%%%%%%%%%%%%%%%%%%%%%%%%%%%%%%%%%%%%%%%%%%%%%%%%%%%%%%%
\begin{abstract}
It is well known that electroencephalograms (EEGs) often contain artifacts due to muscle activity, eye blinks, and various other causes. Detecting such artifacts is an essential first step toward a correct interpretation of EEGs. Although much effort has been devoted to semi-automated and automated artifact detection in EEG, the problem of artifact detection remains challenging. In this paper, we propose a convolutional neural network (CNN) enhanced by transformers using belief matching (BM) loss for automated detection of five types of artifacts: chewing, electrode pop, eye movement, muscle, and shiver. Specifically, we apply these five detectors at individual EEG channels to distinguish artifacts from background EEG. Next, for each of these five types of artifacts, we combine the output of these channel-wise detectors to detect artifacts in multi-channel EEG segments. These segment-level classifiers can detect specific artifacts with a balanced accuracy (BAC) of 0.947, 0.735, 0.826, 0.857, and 0.655 for chewing, electrode pop, eye movement, muscle, and shiver artifacts, respectively. Finally, we combine the outputs of the five segment-level detectors to perform a combined binary classification (any artifact vs.~background). The resulting detector achieves a sensitivity (SEN) of 60.4\%, 51.8\%, and 35.5\%, at a specificity (SPE) of 95\%, 97\%, and 99\%, respectively. This artifact detection module can reject artifact segments while only removing a small fraction of the background EEG, leading to a cleaner EEG for further analysis.
%\newline
%\indent \textit{Clinical relevance}— The artifact detector can help reject 42\% of artifacts while keeping 95\% of the normal EEG signals intact.
\vspace{0.3cm}

\end{abstract}

%%%%%%%%%%%%%%%%%%%%%%%%%%%%%%%%%%%%%%%%%%%%%%%%%%%%%%%%%%%%%%%%%%%%%%%%%%%%%%%%
\section{INTRODUCTION}
Electroencephalography (EEG) is a widely used technology in neurology, e.g., helpful for the diagnosis of epilepsy~\cite{chang2018evaluation}. However, EEG recordings often contain artifacts, which can be due to eyeblinks, head movements, chewing, interference from electronic equipment, and other causes~\cite{jiang2019removal}. These artifacts may resemble epileptiform abnormalities or other transient waveforms, resulting in mistakes during annotation~\cite{pion2019iclabel}. Knowledge of the plausible scalp distribution of EEG abnormalities is essential to distinguish artifacts from brain waves~\cite{abdi2021eeg}. For instance, muscle artifacts usually appear in multiple channels, whereas artifacts such as electrode pop may only be visible in a single channel.

\begin{figure}
\begin{minipage}[b]{0.46\linewidth}
  \centering
  \centerline{\includegraphics[width=4.5cm]{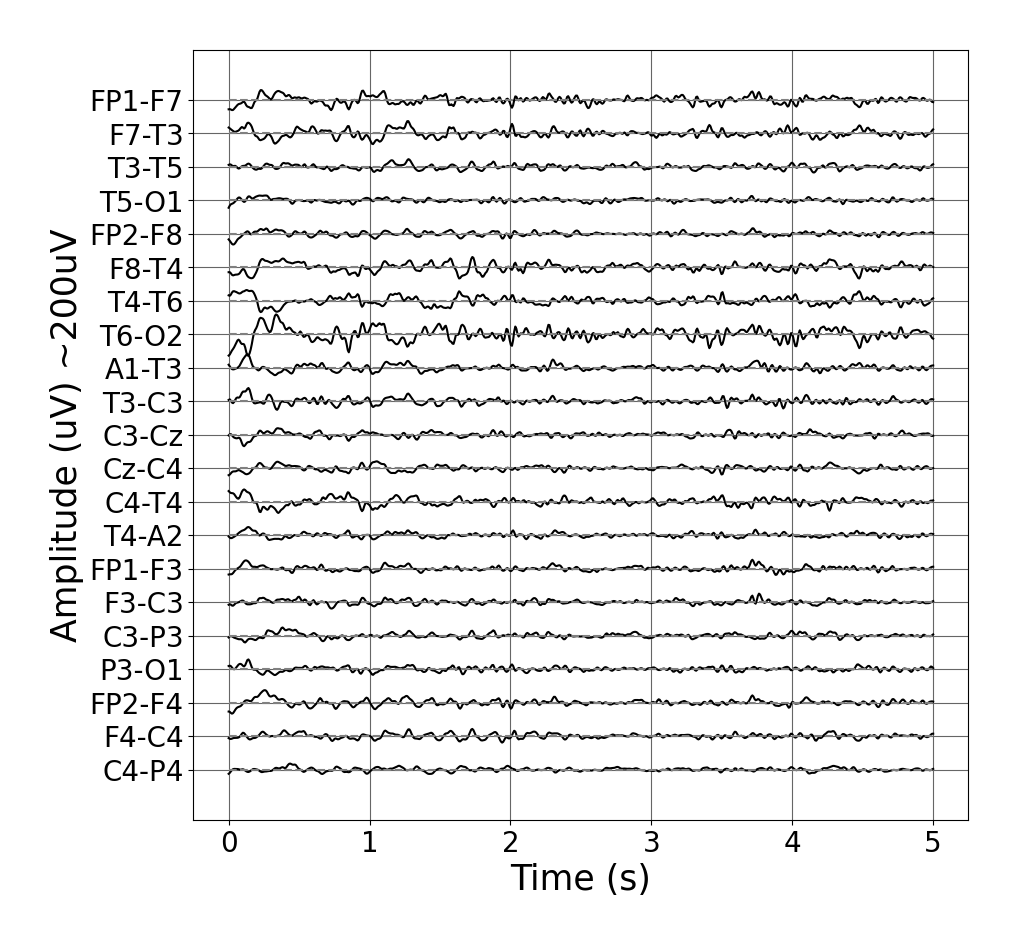}}
  \vspace{-0.3cm}
  \centerline{(a) Background.}\medskip
\end{minipage}
\hfill
\begin{minipage}[b]{0.46\linewidth}
  \centering
  \centerline{\includegraphics[width=4.5cm]{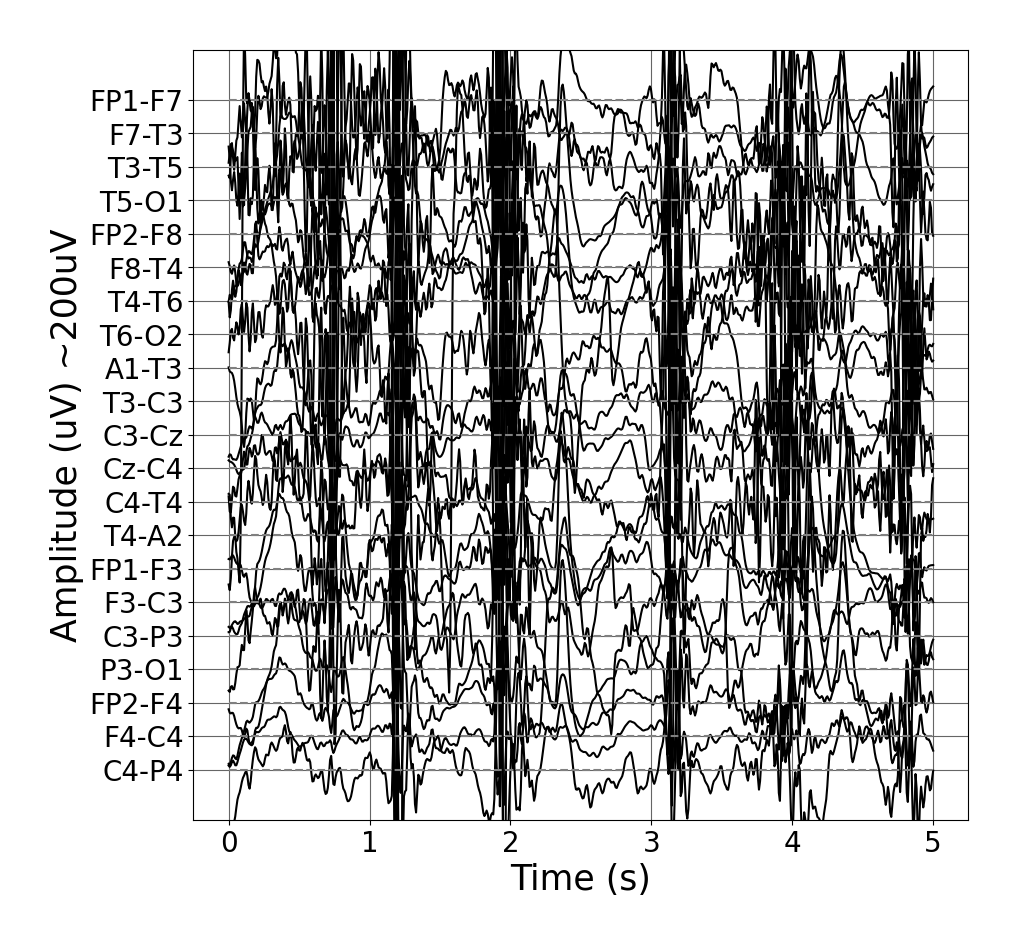}}
  \vspace{-0.3cm}
  \centerline{(b) Chewing.}\medskip
\end{minipage}
\hfill
\begin{minipage}[b]{0.46\linewidth}
  \centering
  \centerline{\includegraphics[width=4.5cm]{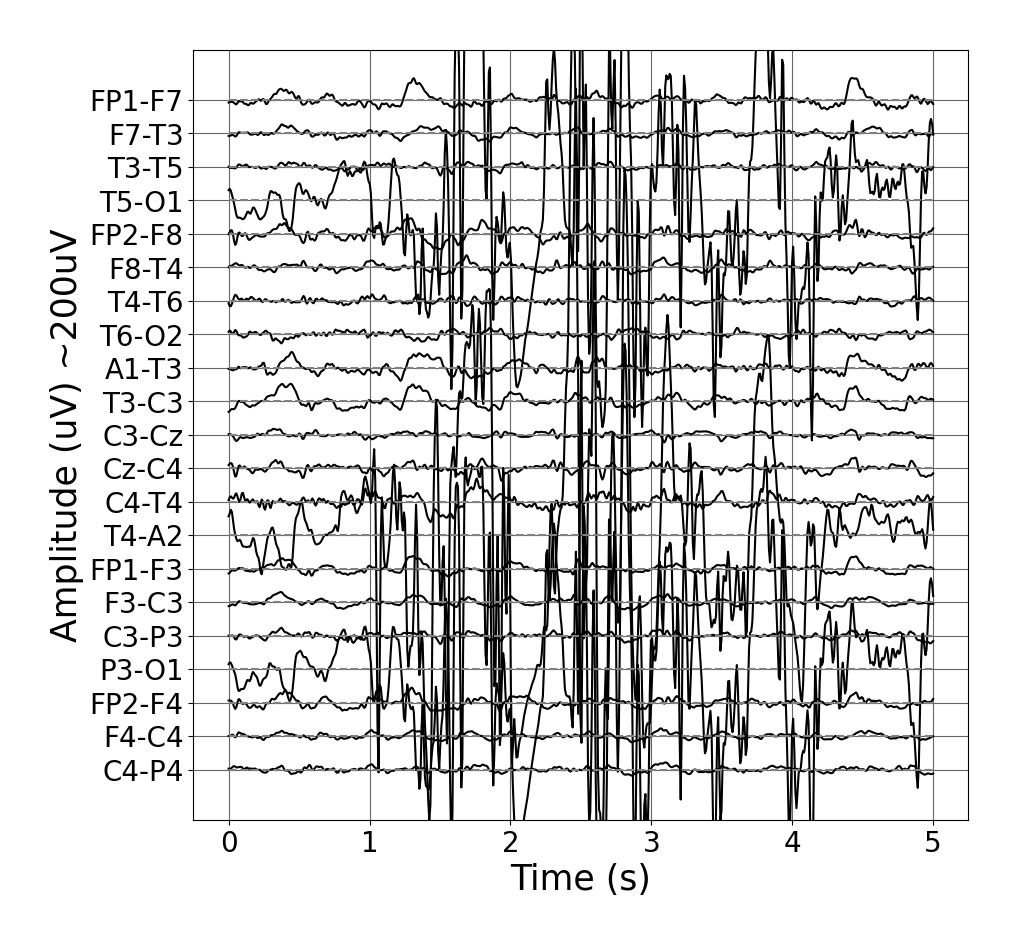}}
  \vspace{-0.3cm}
  \centerline{(c) Electrode pop.}\medskip
\end{minipage}
\hfill
\begin{minipage}[b]{0.46\linewidth}
  \centering
  \centerline{\includegraphics[width=4.5cm]{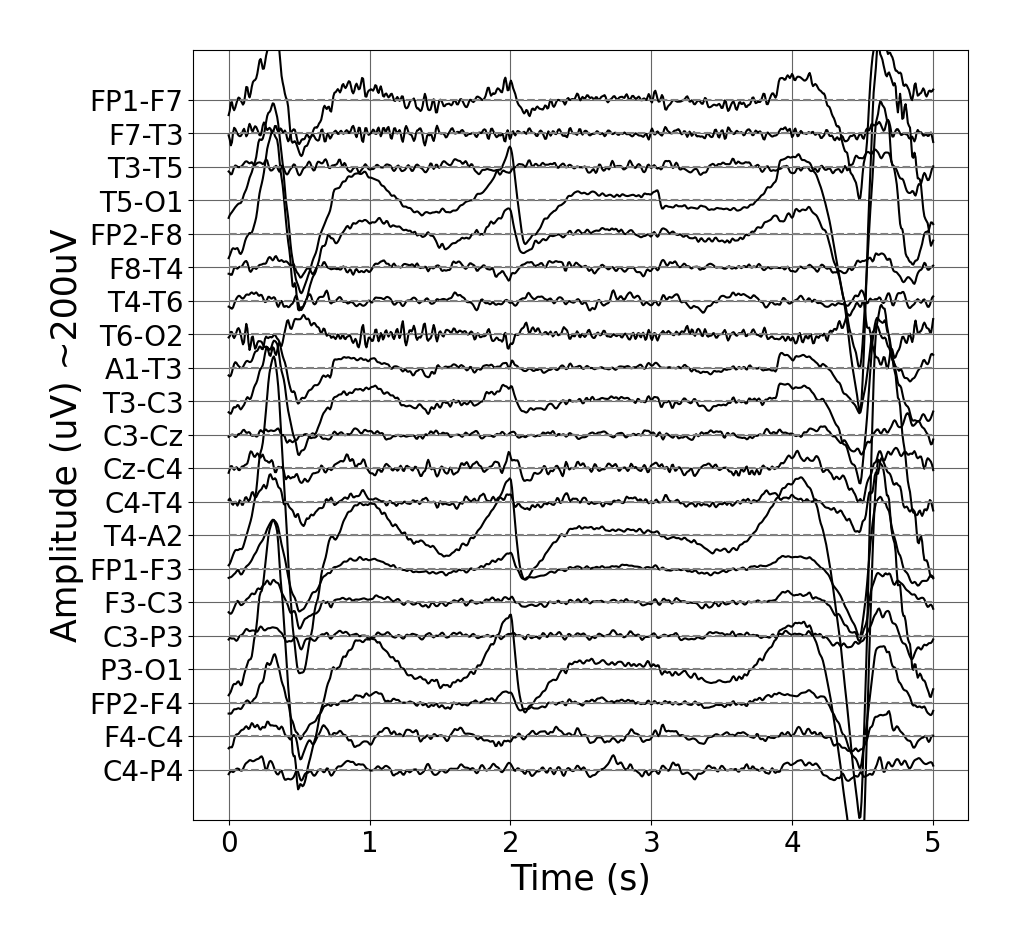}}
  \vspace{-0.3cm}
  \centerline{(d) Eye movement.}\medskip
\end{minipage}
\hfill
\begin{minipage}[b]{0.46\linewidth}
  \centering
  \centerline{\includegraphics[width=4.5cm]{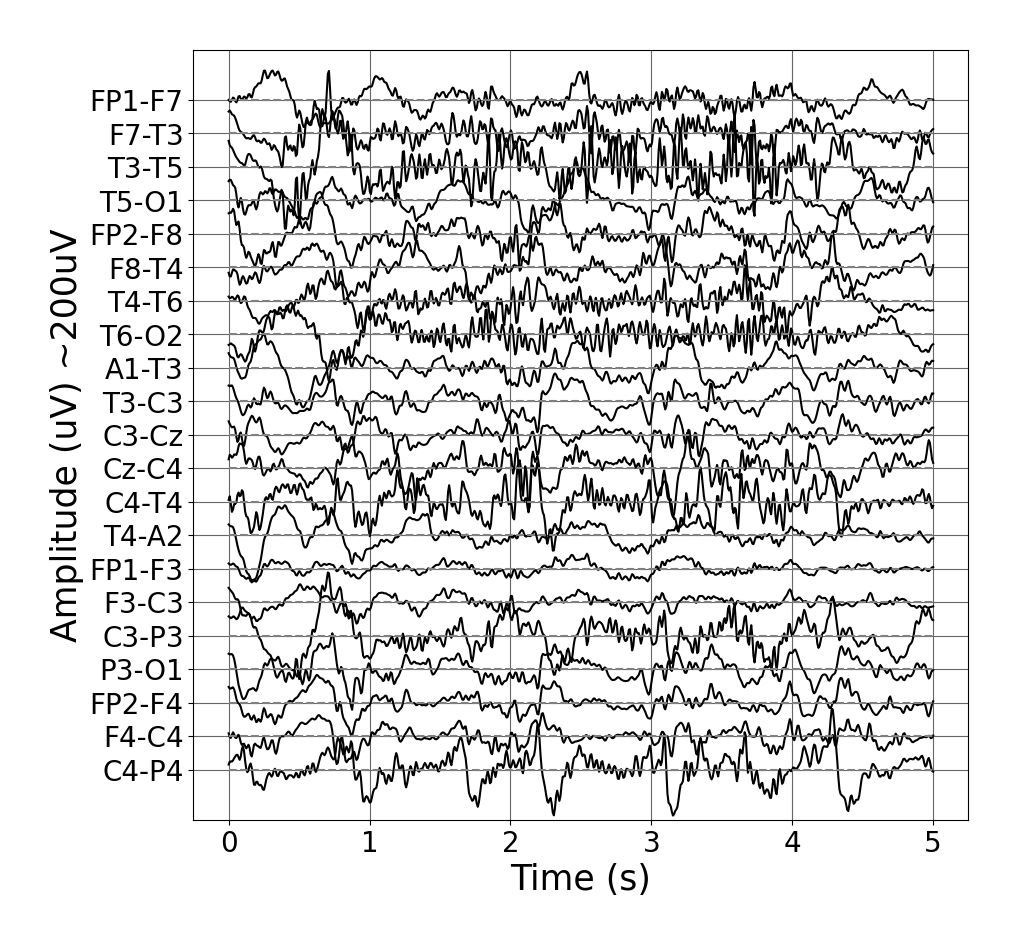}}
  \vspace{-0.3cm}
  \centerline{(e) Muscle.}\medskip
\end{minipage}
\hfill
\begin{minipage}[b]{0.46\linewidth}
  \centering
  \centerline{\includegraphics[width=4.5cm]{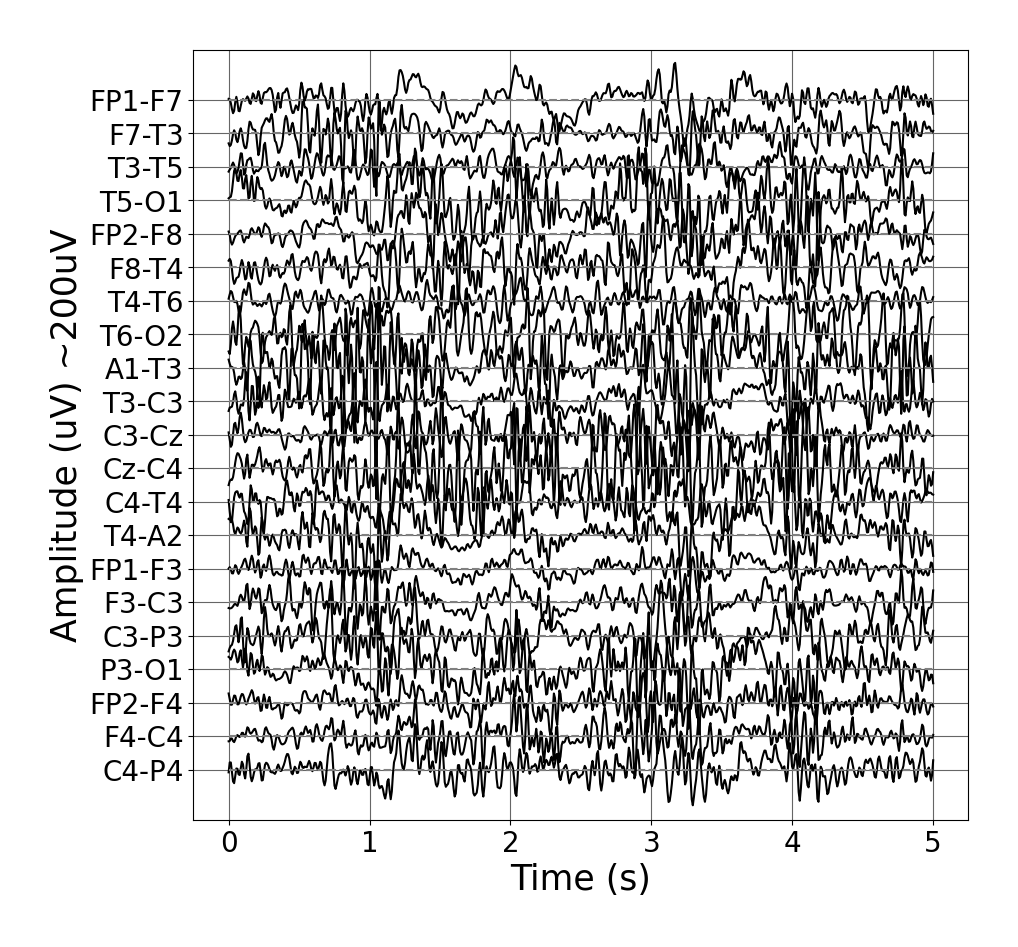}}
  \vspace{-0.3cm}
  \centerline{(f) Shiver.}\medskip
\end{minipage}
\vspace{-0.3cm}
\caption{Examples of clean EEG and EEG with various types of artifacts.}
\label{fig:type_of_artifacts}
\vspace{-0.5cm}

\end{figure}

When reading EEGs, one must distinguish artifacts from brain waves. An automatic artifact detection system can improve the readability of an EEG \cite{jiang2019removal}. Common artifact rejection and detection methods includes {high amplitude rejection~\cite{thomas2021automated}}, {common average referencing (CAR)~\cite{thangavel2021time}}, {independent component analysis (ICA)~\cite{abdi2021eeg, pion2019iclabel}}, {wavelet transforms (WT)~\cite{abdi2021eeg}}, {machine learning~\cite{roy2019machine}}, {convolutional neural networks (CNN)~\cite{mashhadi2020deep, pion2019iclabel}}, and {generative adversarial networks (GAN)~\cite{pion2019iclabel}}. The majority of the studies are not validated on large datasets (more than 100 patients), but instead on small datasets (less than 100 patients) or semi-simulated datasets~\cite{abdi2021eeg, mashhadi2020deep}~by injecting noise into regular EEGs. Additionally, most studies detect artifacts directly from multi-channel segments~\cite{roy2019machine}; as a result, many of those methods are only applicable to a fixed number of channels, whereas the proposed method can be applied to EEG with any number of channels. Ultimately, most studies failed to deploy proper evaluation metrics to measure the effectiveness of their methods~\cite{thomas2021automated, thangavel2021time}. Consequently, it is challenging to compare the existing artifact detectors.

\begin{figure}
\centering
\includegraphics[width=7cm]{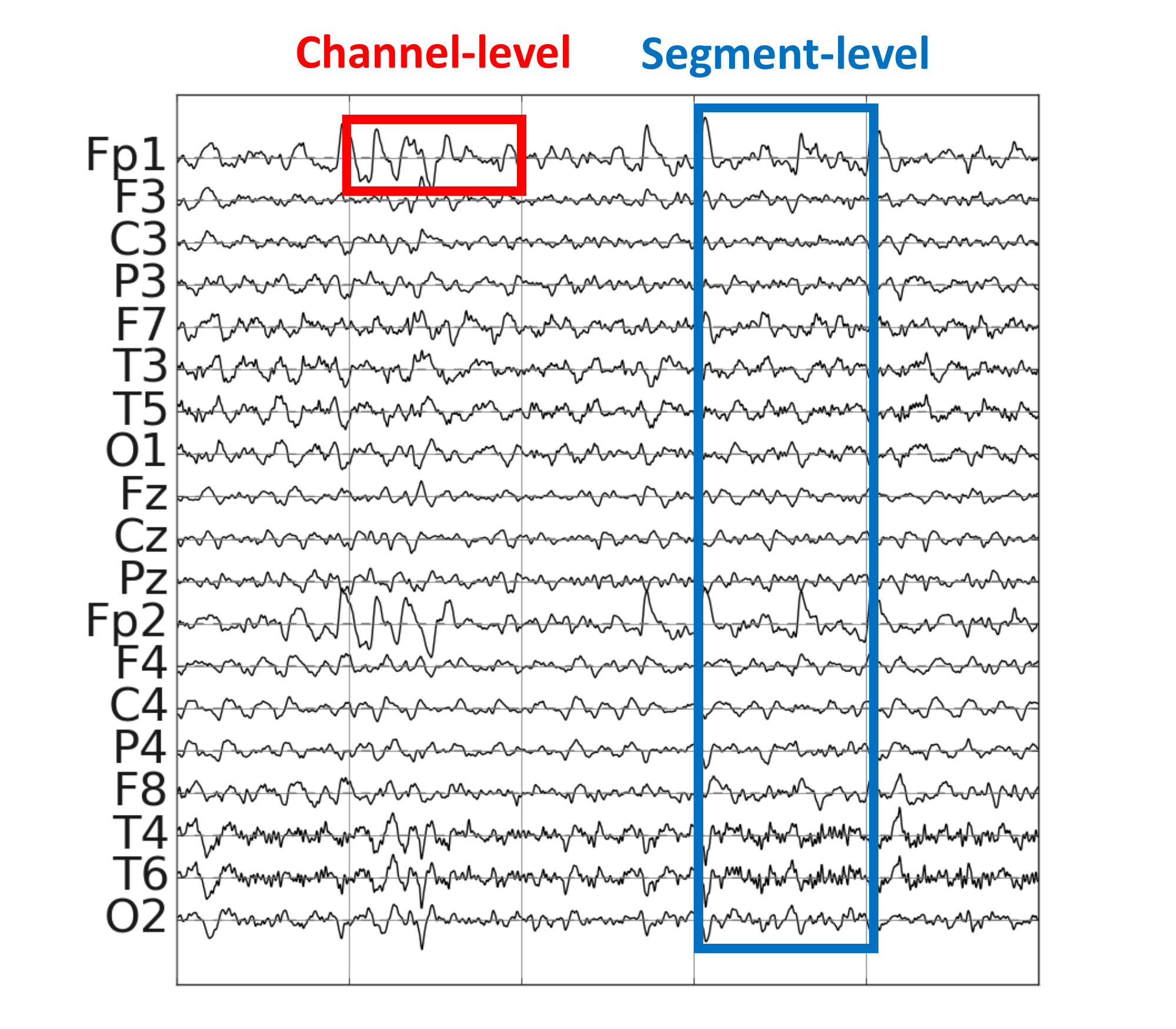}
\vspace*{-0.5cm}
\caption{Channel-level and segment-level analysis of EEG.}
\label{fig:EEG_scale}
\end{figure}

In this paper, we propose a CNN equipped with a transformer (CNN-TRF) trained through a belief matching loss (BM) to detect five different types of artifacts (see Figure~\ref{fig:type_of_artifacts}) from the TUH Artifact (TUH-ART) dataset. The proposed system detects artifacts in individual EEG channels and also in multi-channel EEG segments (see Figure~\ref{fig:EEG_scale})~\cite{peh2021multi}. The artifact detector can detect specific artifacts at segment-level with a balanced accuracy (BAC) of 0.947, 0.735, 0.826, 0.857, and 0.655 for chewing, electrode pop, eye movement, muscle, and shiver artifacts, respectively. When combined to perform binary artifact classification (any artifact type vs. background EEG), the binary artifact detector achieves a sensitivity (SEN) of 42.0\%, 32.0\%, and 13.3\% at 95\%, 97\%, and 99\% specificity (SPE), respectively. This artifact detector can detect specific artifacts and reject them from EEGs, resulting in a cleaner EEG for a better reviewing experience.

%One of the most popular software is {ICLabel~\cite{pion2019iclabel}}, which consists of an EEG independent component classifier, accompanied by its huge dataset. The ICLabel deployed ICA to extract independent components (IC). Next, they deployed different variation of CNNs and GAN to classify the ICs.

\section{Methods}
\subsection{Scalp EEG recordings and preprocessing}
In this study, we analyzed the public TUH Artifact Corpus (TUH-ART), containing EEGs with artifact annotations~\cite{hamid2020temple}. The dataset consists of five artifacts types: chewing, electrode pop, eye movement, muscle, and shiver (see Table~\ref{tab:TUH_ART_dataset}). On each EEG, we applied a Butterworth notch filter ($4^{th}$ order) at 60Hz (USA) to remove interference and a 1Hz high-pass filter ($4^{th}$ order) to remove noise~\cite{peh2021multi}. We downsampled all EEGs to 128Hz. We trained the artifact detectors via 5-fold cross-validation (CV), where each fold contains different patients and similar distribution across all five artifact types.

%standardized the gain of the EEG signals to 1 and 

\begin{table}
\centering
\caption{Summary of the TUH ART EEG dataset.}
\label{tab:TUH_ART_dataset}
\scalebox{0.95}{
\centering
\begin{tabular}{ccc} 
\hline \hline
\textbf{Background/Artifacts} & \textbf{Duration (hr)} & \textbf{Number of Events} \\ 
\hline \hline
Background & 62.533 & - \\
Chewing & 0.995 & 804 \\
Electrode Pop & 5.554 & 6090 \\
Eye Movement & 8.04 & 9480 \\
Muscle & 16.303 & 11267 \\
Shiver & 0.047 & 31 \\
\hline \hline
\vspace{-0.5cm}
\end{tabular}
}
\end{table}

\subsection{Channel-level Artifact Detection}
\begin{figure}
\centering
\hspace*{0.5cm}
\includegraphics[width=8cm]{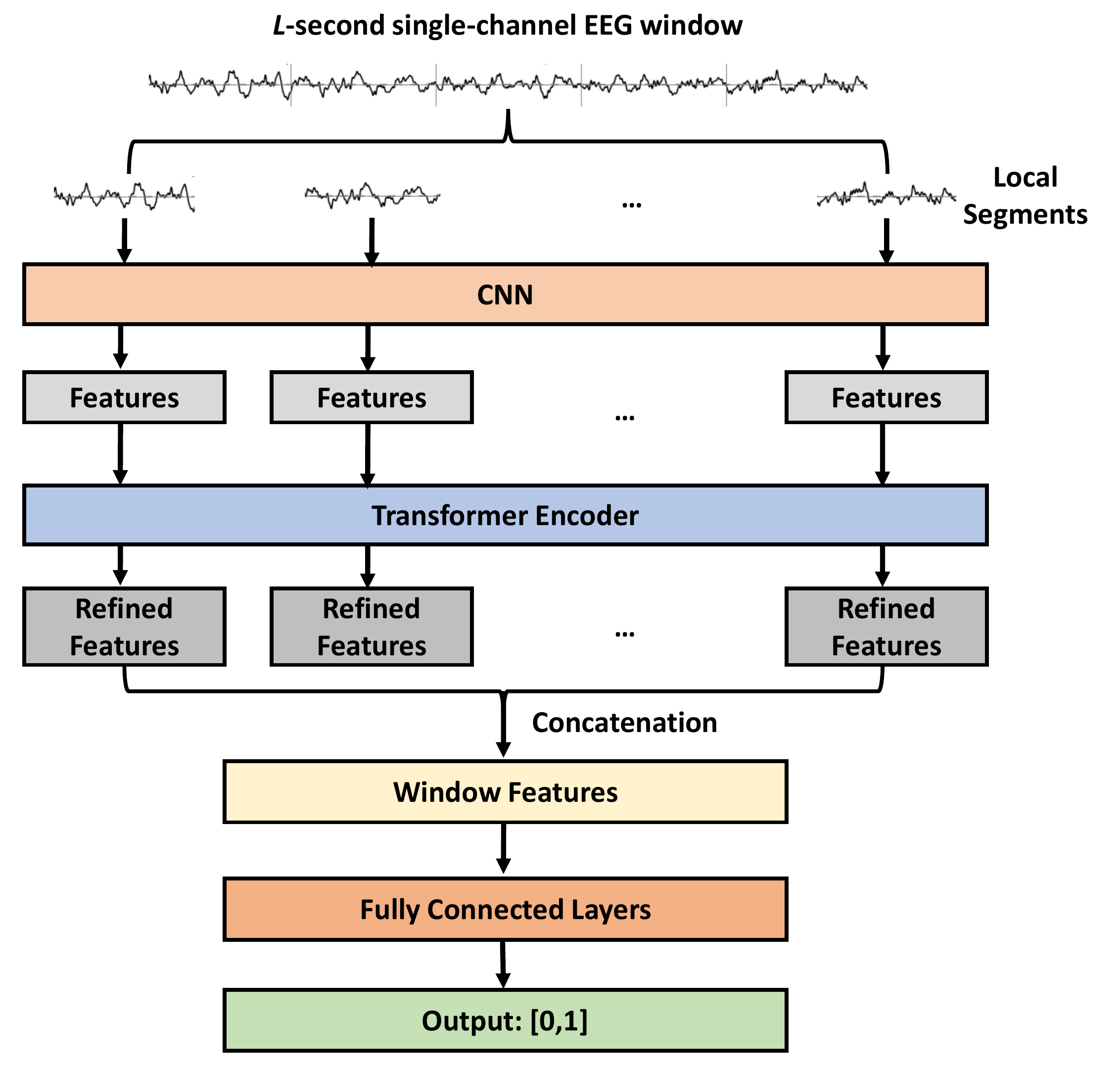}
\caption{CNN with transformer encoder.}
\label{fig:modified_transformer_CNN}
\end{figure}

First, we develop a system to detect artifacts at individual EEG channels (channel-level analysis). We train a separate channel-wise detector on the TUH-ART dataset for each of the five artifact types. The channel-level artifact detector is a CNN cascaded with a transformer, while the learning objective function is a BM loss~\cite{joo2020being} (see Figure~\ref{fig:modified_transformer_CNN}).

A CNN is not adequate for modelling correlations between distant data points. This inherent limitation makes CNN less suitable for time series, especially when correlations over relatively long periods are expected, such as long artifacts patterns (e.g., eye movement artifacts). Therefore, we augment the CNN with a transformer to compensate for this limitation since transformers can extract long-range patterns in the features extracted by the CNN.

In addition, it is essential to have a reliable measure of the uncertainty associated with a detection (output of the neural network) such that we can be confident in the detections with low uncertainty. To this end, we deploy a BM loss instead of the traditional softmax (SM) loss, as it yields more reliable uncertainty estimates~\cite{joo2020being}. The BM framework is a Bayesian approach that views the binary classification from a distribution matching perspective, making it a more reliable detector. Moreover, Joo \textit{et al.} observed improvements in generalization, a desirable property for the application at hand~\cite{joo2020being}. The BM loss is defined as:
\begin{equation} \label{eq:BM_Loss}
\mathcal{L}(\textbf{W}) \approx
-\frac{1}{m} \sum_{i=1}^{m} l_\text{EB} 
\left(y^{(i)}, \alpha^{\textbf{W}}(\textbf{x}^{(i)}) \right),
\end{equation}

\noindent where $\textbf{x}^{(i)}$ and $y^{(i)}$ are the $i$-th training data and its label, respectively, $m$ is the total number of training data, {$l_{\text{EB}}$ is the evidence lower bound (ELBO)~\cite{joo2020being}}, and $\alpha^{\textbf{W}}=\text{exp} (\textbf{W})$, where $\textbf{W}$ are the weights of the neural network classifier.

The input of the CNN-TRF is the $L$-second single-channel EEG window that is split into 0.5s local segments with 25\% overlap (see Figure~\ref{fig:modified_transformer_CNN}). We trained the model with different window lengths $L$, i.e., 1, 3, and 5s. We varied the window lengths to determine the best window length to detect artifacts. For instance, a short window length of 1s is adequate to remove short-duration eye blinks, while a long window length of 5s is suitable to remove long-duration muscle artifacts.

The CNN architecture consists of 5 convolution layers, all with a filter size of 3 and a ReLU activation function. The number of filters is 8, 16, 32, 64, and 128 in layers 1, 2, 3, 4, and 5, respectively. After each convolution layer, we apply max-pooling with stride 2. Next, we deploy a transformer encoder to identify patterns in the features extracted by the CNN~\cite{vaswani2017attention}. The encoder relies on an activation function that maps the query and a set of key-value pairs to an output. Here, the local features extracted by the CNN are the query, key, and value simultaneously. We set the number of heads in the transformer to the commonly chosen value of {8~\cite{vaswani2017attention}} and the number of neurons in the hidden layer of the feed-forward network (FNN) module to 1024. Two fully connected (FC) layers containing 100 and 2 neurons follow the CNN-TRF module. Before the final FC layer, we include a dropout layer with a probability of 0.5. The output of the second FC is the prediction for that particular window of EEG. Finally, we applied the Adam optimizer~\cite{kingma2014adam} with an initial learning rate equal to $10^{-4}$ to minimize the BM loss. The batch size for training is 1000. We applied balanced training to avoid overfitting during training by applying weights to each class. We optimized the hyperparameters of the CNN-TRF via nested CV on the training data with an 80\%:20\% split for training and validation.

\subsection{Segment-level Artifact Detection}
\begin{figure*}
\centering
\includegraphics[width=16cm]{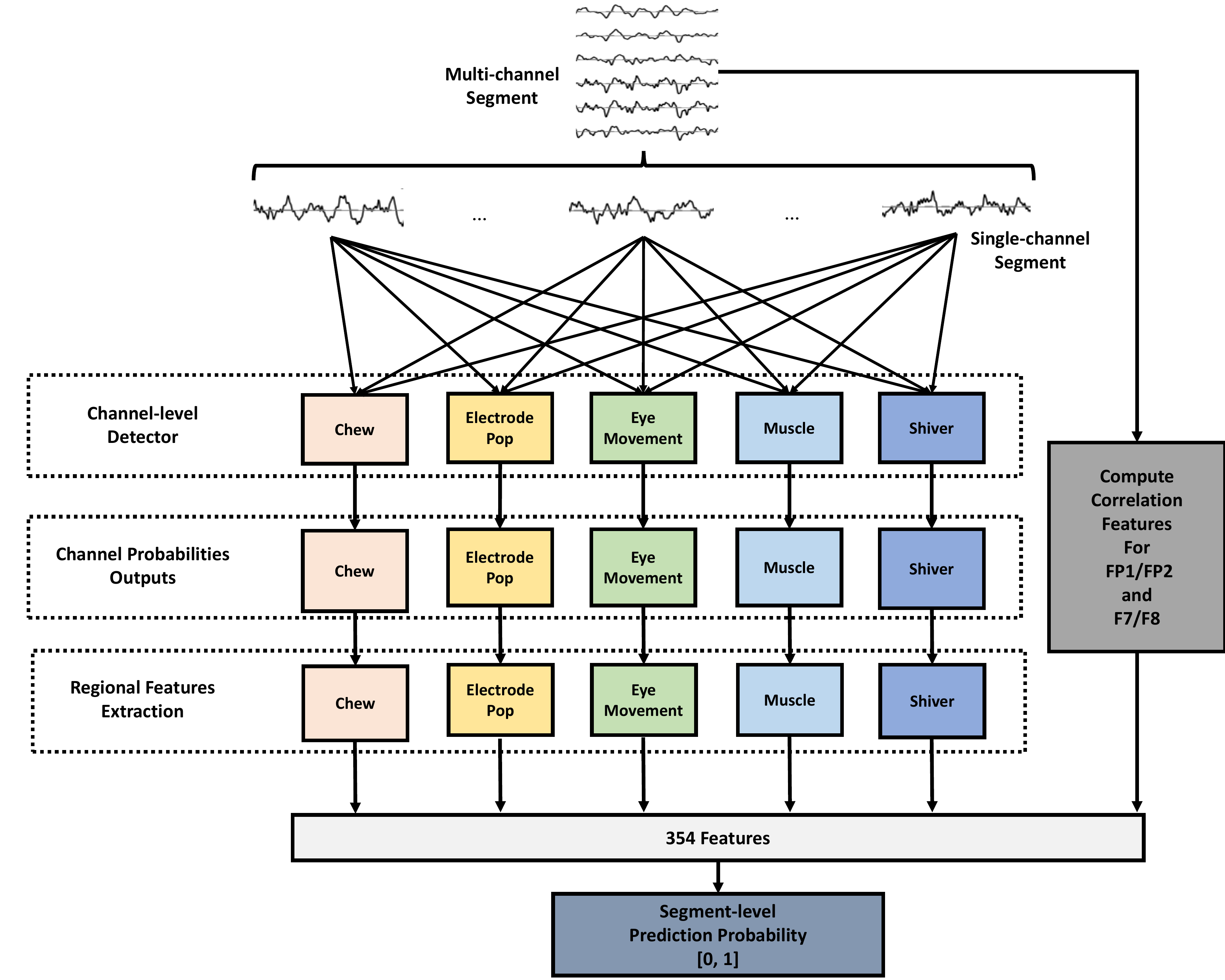}
\caption{Artifact detection pipeline for segment-level classification.}
\vspace{-0.5cm}
\label{fig:artifact_detector}
\end{figure*}

Next, we wish to detect artifacts in multi-channel segments (see Figure~\ref{fig:artifact_detector}). To perform binary classification of a specific artifact type, we performed the following:

\begin{enumerate}

  \item Perform channel-level predictions on all channels in a multi-channel segment.
  
  \item With the set of probabilities outputs and knowledge of their location, we distribute probability outputs accordingly to seven regions: frontal, frontal-temporal, non-frontal (all non-frontal channels), central, parietal, occipital, and the entire scalp.
  
  \item From each region, we extract statistical features: mean, median, standard deviation, maximum values, minimum values, and the histogram features (5 bins, range: [0,1]). This corresponds to 10 features per region.
  
\end{enumerate}

This results in $70$ features for each artifact type. Additionally, we include the cross-correlation and auto-correlation of the signals between channel FP1/FP2 and channel F7/F8 to account for eye blink features in each feature set. Eventually, we obtain $74$ features from each multi-channel segment for the training and testing of each artifact class. We performed segment-level binary classification (specific artifact vs. background) for each artifact type. We classify the features with {CatBoost~\cite{dorogush2018catboost}}, and optimize the hyperparameters by grid search. 

Lastly, we concatenate the probability outputs from the five segment-level classifiers. With all the features, we trained a CatBoost classifier to detect any of the five artifacts types; in other words, this system is designed to determine whether a multi-channel EEG segment is clean or contains artifact(s). Finally, we evaluate the systems with the following metrics: area under the receiver operator characteristic (AUC), area under the precision-recall curve (AUPRC), accuracy (ACC), balanced accuracy (BAC), sensitivity (SEN), and specificity (SPE)~\cite{peh2021multi}. NVIDIA GeForce GTX1080 GPU machines, Keras 2.2.0 and TensorFlow 2.6.0 were adopted in this study.

%Additionally, for the channel-level artifact detector, we deployed the expected calibration error to measure the calibration error and behavior of the CNN-TRF model (ECE)~\cite{guo2017calibration}.

\section{Results}

\subsection{Channel-level Artifact Detection Results}
We displayed the results for the channel-level artifact detector in Table~\ref{tab:channel_artifact_results}. The chewing artifact achieved the best BAC (0.911), while the electrode pop artifact achieved the poorest BAC (0.716). For most artifacts, the best BAC is achieved at a window length of 5s. However, the best BAC for each artifact class is obtained at a different window length.

\begin{table}[htp]
\centering
\caption{Channel-level artifact detection results.}
\label{tab:channel_artifact_results}
\scalebox{0.8}{
\centering
\begin{tabular}{cccccccc} 

\hline \hline
\multirow{2}{*}{\begin{tabular}[c]{@{}c@{}}\textbf{Artifact}\\\textbf{Type}\end{tabular}} & \multirow{2}{*}{\begin{tabular}[c]{@{}c@{}}\textbf{Window}\\\textbf{Length}\end{tabular}} & \multicolumn{6}{c}{\textbf{Binary}} \\ 
\cline{3-8}
 & & \textbf{AUC} & \textbf{AUPRC} & \textbf{ACC} & \textbf{BAC} & \textbf{SEN} & \textbf{SPE} \\ 
\hline \hline

\multirow{3}{*}{Chewing} & 1 & 0.961 & 0.950 & 0.901 & 0.901 & 0.894 & 0.907 \\
 & 3 & \cellcolor[gray]{0.9}{0.966} & \cellcolor[gray]{0.9}{0.904} & \cellcolor[gray]{0.9}{0.933} & \cellcolor[gray]{0.9}{0.911} & \cellcolor[gray]{0.9}{0.856} & \cellcolor[gray]{0.9}{0.966} \\
 & 5 & 0.967 & 0.864 & 0.950 & 0.906 & 0.835 & 0.978 \\ 
\hline

\multirow{3}{*}{Electrode pop} & 1 & 0.792 & 0.926 & 0.802 & 0.663 & 0.914 & 0.412 \\
 & 3 & \cellcolor[gray]{0.9}{0.802} & \cellcolor[gray]{0.9}{0.802} & \cellcolor[gray]{0.9}{0.709} & \cellcolor[gray]{0.9}{0.716} & \cellcolor[gray]{0.9}{0.718} & \cellcolor[gray]{0.9}{0.713} \\
 & 5 & 0.783 & 0.681 & 0.711 & 0.684 & 0.559 & 0.809 \\ 
\hline

\multirow{3}{*}{Eye movement} & 1 & 0.856 & 0.921 & 0.795 & 0.734 & 0.905 & 0.564 \\
 & 3 & 0.866 & 0.799 & 0.807 & 0.788 & 0.723 & 0.853 \\
 & 5 & \cellcolor[gray]{0.9}{0.895} & \cellcolor[gray]{0.9}{0.758} & \cellcolor[gray]{0.9}{0.877} & \cellcolor[gray]{0.9}{0.792} & \cellcolor[gray]{0.9}{0.644} & \cellcolor[gray]{0.9}{0.940} \\ 
\hline

\multirow{3}{*}{Muscle} & 1 & 0.794 & 0.949 & 0.936 & 0.760 & 0.990 & 0.529 \\
 & 3 & 0.931 & 0.973 & 0.907 & 0.836 & 0.961 & 0.711 \\
 & 5 & \cellcolor[gray]{0.9}{0.934} & \cellcolor[gray]{0.9}{0.957} & \cellcolor[gray]{0.9}{0.888} & \cellcolor[gray]{0.9}{0.861} & \cellcolor[gray]{0.9}{0.942} & \cellcolor[gray]{0.9}{0.780} \\ 
\hline

\multirow{3}{*}{Shiver} & 1 & 0.657 & 0.138 & 0.948 & 0.527 & 0.091 & 0.994 \\
 & 3 & 0.610 & 0.056 & 0.986 & 0.563 & 0.182 & 0.943 \\
 & 5 & \cellcolor[gray]{0.9}{0.756} & \cellcolor[gray]{0.9}{0.066} & \cellcolor[gray]{0.9}{0.994} & \cellcolor[gray]{0.9}{0.621} & \cellcolor[gray]{0.9}{0.364} & \cellcolor[gray]{0.9}{0.878} \\ 
\hline \hline

\end{tabular}
}
\end{table}

\subsection{Segment-level Artifact Detection Results: Binary}
We displayed the results for the binary artifact classification for each artifact class in Table~\ref{tab:segment_artifact_results_binary}. We performed classification using two feature sets. The first feature set (specific features) deployed only the features extracted from its respective channel-level artifact detector. The second feature set (all features) utilized the features extracted from all five channel-level artifact detectors. Similarly, the chewing artifact achieved the best BAC, while the electrode pop artifact achieved the poorest BAC. We noticed that the best BAC is achieved at different window lengths with different feature sets for each artifact type. Generally, the best results for the electrode pop and eye movement artifacts are obtained at shorter window lengths. In contrast, chewing, muscle, and shiver artifacts achieve better results at longer window lengths.

\begin{table}[htp]
\centering
\caption{Segment-level binary artifact detection results.}
\label{tab:segment_artifact_results_binary}
\scalebox{0.8}{
\centering
\begin{tabular}{cccccccc} 

\hline \hline
\multirow{2}{*}{\begin{tabular}[c]{@{}c@{}}\textbf{Artifact}\\\textbf{Type}\end{tabular}} & \multirow{2}{*}{\begin{tabular}[c]{@{}c@{}}\textbf{Window}\\\textbf{Length}\end{tabular}} & \multicolumn{6}{c}{\textbf{Binary}} \\ 
\cline{3-8}
 & & \textbf{AUC} & \textbf{AUPRC} & \textbf{ACC} & \textbf{BAC} & \textbf{SEN} & \textbf{SPE} \\ 
\hline \hline

\multirow{3}{*}{Chewing} & 1 & 0.953 & 0.730 & 0.971 & 0.922 & 0.868 & 0.976 \\
 & 3 & \cellcolor[gray]{0.9}{0.963} & \cellcolor[gray]{0.9}{0.742} & \cellcolor[gray]{0.9}{0.983} & \cellcolor[gray]{0.9}{0.941} & \cellcolor[gray]{0.9}{0.895} & \cellcolor[gray]{0.9}{0.987} \\
 & 5 & 0.936 & 0.734 & 0.986 & 0.929 & 0.866 & 0.992 \\ 
\hline

\multirow{3}{*}{Electrode pop} & 1 & \cellcolor[gray]{0.9}{0.815} & \cellcolor[gray]{0.9}{0.362} & \cellcolor[gray]{0.9}{0.814} & \cellcolor[gray]{0.9}{0.756} & \cellcolor[gray]{0.9}{0.675} & \cellcolor[gray]{0.9}{0.837} \\
 & 3 & 0.824 & 0.264 & 0.850 & 0.739 & 0.601 & 0.877 \\
 & 5 & 0.813 & 0.218 & 0.878 & 0.725 & 0.551 & 0.899 \\ 
\hline

\multirow{3}{*}{Eye movement} & 1 & \cellcolor[gray]{0.9}{0.895} & \cellcolor[gray]{0.9}{0.609} & \cellcolor[gray]{0.9}{0.823} & \cellcolor[gray]{0.9}{0.819} & \cellcolor[gray]{0.9}{0.816} & \cellcolor[gray]{0.9}{0.823} \\
 & 3 & 0.881 & 0.424 & 0.870 & 0.814 & 0.745 & 0.883 \\
 & 5 & 0.876 & 0.374 & 0.881 & 0.824 & 0.759 & 0.889 \\ 
\hline

\multirow{3}{*}{Muscle} & 1 & 0.894 & 0.736 & 0.796 & 0.832 & 0.939 & 0.725 \\
 & 3 & \cellcolor[gray]{0.9}{0.903} & \cellcolor[gray]{0.9}{0.659} & \cellcolor[gray]{0.9}{0.815} & \cellcolor[gray]{0.9}{0.858} & \cellcolor[gray]{0.9}{0.940} & \cellcolor[gray]{0.9}{0.775} \\
 & 5 & 0.902 & 0.579 & 0.814 & 0.856 & 0.922 & 0.791 \\ 
\hline

\multirow{3}{*}{Shiver} & 1 & 0.691 & 0.027 & 0.993 & 0.516 & 0.034 & 0.997 \\
 & 3 & 0.770 & 0.068 & 0.994 & 0.530 & 0.062 & 0.998 \\
 & 5 & \cellcolor[gray]{0.9}{0.661} & \cellcolor[gray]{0.9}{0.308} & \cellcolor[gray]{0.9}{0.996} & \cellcolor[gray]{0.9}{0.655} & \cellcolor[gray]{0.9}{0.311} & \cellcolor[gray]{0.9}{0.997} \\ 
\hline \hline

\end{tabular}
}
\end{table}

\subsection{Segment-level Artifact Detection Results: Multi-Class}
\begin{figure*}[htp]
\begin{minipage}[b]{0.32\linewidth}
  \centering
  \centerline{\includegraphics[width=7cm]{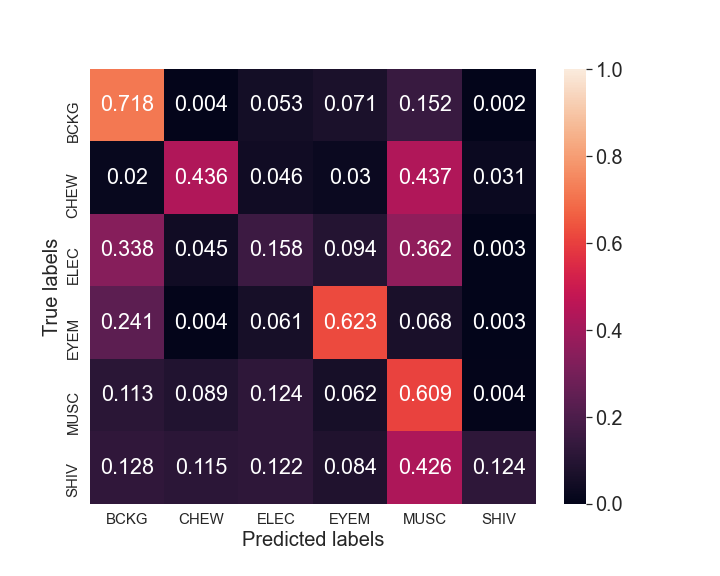}}
  \centerline{(a) Window length of 1s.}\medskip
\end{minipage}
\hfill
\begin{minipage}[b]{0.32\linewidth}
  \centering
  \centerline{\includegraphics[width=7cm]{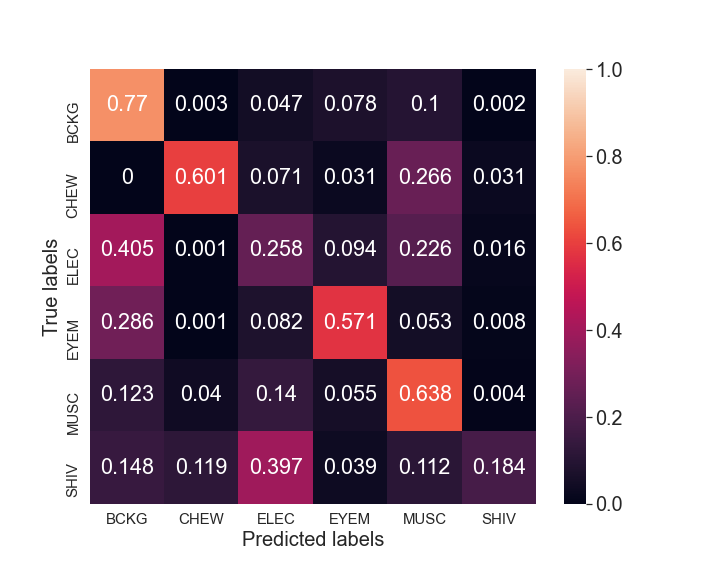}}
  \centerline{(b) Window length of 3s.}\medskip
\end{minipage}
\hfill
\begin{minipage}[b]{0.32\linewidth}
  \centering
  \centerline{\includegraphics[width=7cm]{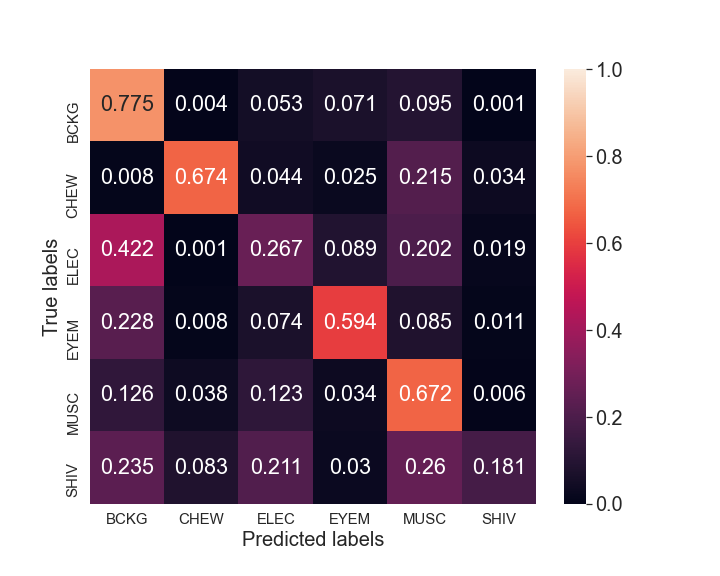}}
  \centerline{(c) Window length of 5s.}\medskip
\end{minipage}
\caption{Confusion matrix of the multi-class segment-level classification at a window length of (a) 1s, (b) 3s, and (c) 5s. The background, chewing, electrode pop, eye movement, muscle, and shiver classes are denoted as BCKG, CHEW, ELEC, EYEM, MUSC, and SHIV, respectively. Note that the confusion matrix is computed by taking the sum of all the predictions, while the results reported in Table~\ref{tab:segment_artifact_results_multi_class} are computed by taking the average of the results obtained from each fold (5-fold CV).}
\label{fig:artifact_confusion}
\end{figure*}

We displayed the results for the multi-class artifact detector in Table~\ref{tab:segment_artifact_results_multi_class}. Additionally, we illustrated the multi-class confusion matrix in Figure~\ref{fig:artifact_confusion}. We split the classification results into individual artifacts to determine performance for each artifact class. Here, the chewing artifact achieved the best BAC, while the shiver artifact achieved the poorest BAC. This is because the shiver artifact is the minority class, with only a small amount of events. Consequently, the performance of the shiver artifact would be poor; in fact, the system rarely predicts any shiver class. Here, the best results for eye movement artifacts are achieved at shorter window lengths, while chewing, electrode pop, muscle, and shiver artifacts achieved better results at longer window lengths. The results achieved here are much poorer than those from the binary classification, as it is more complicated to distinguish multiple labels simultaneously.

\begin{table}[htp]
\centering
\caption{Segment-level multi-class artifact detection results.}
\label{tab:segment_artifact_results_multi_class}
\scalebox{0.8}{
\centering
\begin{tabular}{cccccccc} 

\hline \hline
\multirow{2}{*}{\begin{tabular}[c]{@{}c@{}}\textbf{Artifact}\\\textbf{Type}\end{tabular}} & \multirow{2}{*}{\begin{tabular}[c]{@{}c@{}}\textbf{Window}\\\textbf{Length}\end{tabular}} & \multicolumn{6}{c}{\textbf{Multi-Class}} \\ 
\cline{3-8}
 & & \textbf{AUC} & \textbf{AUPRC} & \textbf{ACC} & \textbf{BAC} & \textbf{SEN} & \textbf{SPE} \\ 
\hline \hline

\multirow{3}{*}{Chewing} & 1 & 0.799 & 0.264 & 0.632 & 0.710 & 0.447 & 0.972 \\
 & 3 & 0.911 & 0.470 & 0.698 & 0.813 & 0.636 & 0.990 \\
 & 5 & \cellcolor[gray]{0.9}{0.939} & \cellcolor[gray]{0.9}{0.629} & \cellcolor[gray]{0.9}{0.725} & \cellcolor[gray]{0.9}{0.879} & \cellcolor[gray]{0.9}{0.768} & \cellcolor[gray]{0.9}{0.990} \\ 
\hline

\multirow{3}{*}{Electrode pop} & 1 & 0.600 & 0.098 & 0.632 & 0.561 & 0.203 & 0.920 \\
 & 3 & 0.645 & 0.098 & 0.698 & 0.586 & 0.248 & 0.924 \\
 & 5 & \cellcolor[gray]{0.9}{0.649} & \cellcolor[gray]{0.9}{0.087} & \cellcolor[gray]{0.9}{0.725} & \cellcolor[gray]{0.9}{0.615} & \cellcolor[gray]{0.9}{0.301} & \cellcolor[gray]{0.9}{0.929} \\ 
\hline

\multirow{3}{*}{Eye movement} & 1 & \cellcolor[gray]{0.9}{0.809} & \cellcolor[gray]{0.9}{0.428} & \cellcolor[gray]{0.9}{0.632} & \cellcolor[gray]{0.9}{0.774} & \cellcolor[gray]{0.9}{0.614} & \cellcolor[gray]{0.9}{0.933} \\
 & 3 & 0.771 & 0.299 & 0.698 & 0.754 & 0.584 & 0.924 \\
 & 5 & 0.775 & 0.258 & 0.725 & 0.764 & 0.596 & 0.931 \\ 
\hline

\multirow{3}{*}{Muscle} & 1 & 0.779 & 0.505 & 0.632 & 0.722 & 0.625 & 0.820 \\
 & 3 & 0.745 & 0.476 & 0.698 & 0.770 & 0.651 & 0.889 \\
 & 5 & \cellcolor[gray]{0.9}{0.755} & \cellcolor[gray]{0.9}{0.448} & \cellcolor[gray]{0.9}{0.725} & \cellcolor[gray]{0.9}{0.789} & \cellcolor[gray]{0.9}{0.682} & \cellcolor[gray]{0.9}{0.895} \\ 
\hline

\multirow{3}{*}{Shiver} & 1 & 0.866 & 0.020 & 0.632 & 0.562 & 0.124 & 0.999 \\
 & 3 & 0.870 & 0.258 & 0.698 & 0.592 & 0.184 & 1.000 \\
 & 5 & \cellcolor[gray]{0.9}{0.575} & \cellcolor[gray]{0.9}{0.007} & \cellcolor[gray]{0.9}{0.725} & \cellcolor[gray]{0.9}{0.590} & \cellcolor[gray]{0.9}{0.181} & \cellcolor[gray]{0.9}{1.000} \\ 
\hline \hline

\end{tabular}
}
\end{table}

\subsection{Segment-level Artifact Detection Results: Multi-Class Multi-Label (MCML)}
We displayed the results for the multi-class multi-label (MCML) segment-level artifact detector in Table~\ref{tab:segment_artifact_results_MCML}. While the outputs are multi-labels, we split the classification results into individual artifacts to determine the most prominent artifact. The chewing artifact achieved the best BAC, while the electrode pop artifact attained the poorest BAC. Overall, better results for eye movement artifacts are obtained at shorter window lengths, while the other artifacts achieved better results at longer window lengths.

\begin{table}[htp]
\centering
\caption{Segment-level multi-class multi-label (MCML) artifact detection results.}
\label{tab:segment_artifact_results_MCML}
\scalebox{0.8}{
\centering
\begin{tabular}{cccccccc} 

\hline \hline
\multirow{2}{*}{\begin{tabular}[c]{@{}c@{}}\textbf{Artifact}\\\textbf{Type}\end{tabular}} & \multirow{2}{*}{\begin{tabular}[c]{@{}c@{}}\textbf{Window}\\\textbf{Length}\end{tabular}} & \multicolumn{6}{c}{\textbf{Multi-Class Multi-Label}} \\ 
\cline{3-8}
 & & \textbf{AUC} & \textbf{AUPRC} & \textbf{ACC} & \textbf{BAC} & \textbf{SEN} & \textbf{SPE} \\ 
\hline \hline

\multirow{3}{*}{Chewing} & 1 & 0.909 & 0.342 & 0.941 & 0.799 & 0.647 & 0.951 \\
 & 3 & 0.905 & 0.480 & 0.967 & 0.846 & 0.715 & 0.978 \\
 & 5 & \cellcolor[gray]{0.9}{0.932} & \cellcolor[gray]{0.9}{0.612} & \cellcolor[gray]{0.9}{0.980} & \cellcolor[gray]{0.9}{0.879} & \cellcolor[gray]{0.9}{0.770} & \cellcolor[gray]{0.9}{0.988} \\ 
\hline

\multirow{3}{*}{Electrode pop} & 1 & 0.690 & 0.168 & 0.843 & 0.610 & 0.327 & 0.893 \\
 & 3 & \cellcolor[gray]{0.9}{0.747} & \cellcolor[gray]{0.9}{0.199} & \cellcolor[gray]{0.9}{0.859} & \cellcolor[gray]{0.9}{0.652} & \cellcolor[gray]{0.9}{0.410} & \cellcolor[gray]{0.9}{0.894} \\
 & 5 & 0.732 & 0.155 & 0.887 & 0.632 & 0.344 & 0.919 \\ 
\hline

\multirow{3}{*}{Eye movement} & 1 & \cellcolor[gray]{0.9}{0.865} & \cellcolor[gray]{0.9}{0.560} & \cellcolor[gray]{0.9}{0.803} & \cellcolor[gray]{0.9}{0.795} & \cellcolor[gray]{0.9}{0.783} & \cellcolor[gray]{0.9}{0.808} \\
 & 3 & 0.856 & 0.398 & 0.850 & 0.769 & 0.666 & 0.872 \\
 & 5 & 0.853 & 0.345 & 0.852 & 0.782 & 0.698 & 0.866 \\ 
\hline

\multirow{3}{*}{Muscle} & 1 & 0.833 & 0.597 & 0.727 & 0.780 & 0.911 & 0.648 \\
 & 3 & \cellcolor[gray]{0.9}{0.859} & \cellcolor[gray]{0.9}{0.564} & \cellcolor[gray]{0.9}{0.777} & \cellcolor[gray]{0.9}{0.811} & \cellcolor[gray]{0.9}{0.880} & \cellcolor[gray]{0.9}{0.743} \\
 & 5 & 0.870 & 0.535 & 0.770 & 0.811 & 0.881 & 0.741 \\ 
\hline

\multirow{3}{*}{Shiver} & 1 & 0.498 & 0.009 & 0.997 & 0.700 & 0.000 & 0.999 \\
 & 3 & 0.769 & 0.021 & 0.995 & 0.798 & 0.000 & 0.998 \\
 & 5 & \cellcolor[gray]{0.9}{0.831} & \cellcolor[gray]{0.9}{0.030} & \cellcolor[gray]{0.9}{0.998} & \cellcolor[gray]{0.9}{0.799} & \cellcolor[gray]{0.9}{0.000} & \cellcolor[gray]{0.9}{0.999} \\ 
\hline \hline

\end{tabular}
}
\end{table}

\subsection{Segment-level Artifact Detection Results: Combined Binary}
Finally, we combined the five segment-level artifact detectors and obtained the results for the combined binary segment-level artifact detector in Table~\ref{tab:segment_artifact_results_binary_all}. We combined all the artifact classes into a single class to distinguish artifacts against backgrounds. At an SPE of 95\%, 97\%, and 99\%, the highest SEN achieved is 0.604, 0.518, and 0.353, respectively, at a window length of 3, 3, and 5, respectively. However, we note that this evaluation approach no longer accounts for the artifact class imbalance; there are much more muscle artifacts than eye movement artifacts. Hence the combined binary segment-level detector will be overfitted to muscle artifacts.

\begin{table}[htp]
\centering
\caption{Segment-level combined binary artifact detection results.}
\label{tab:segment_artifact_results_binary_all}
\scalebox{0.8}{
\centering
\begin{tabular}{ccccccccc} 
\hline \hline
\multirow{2}{*}{\begin{tabular}[c]{@{}c@{}}\textbf{Window}\\\textbf{Length}\end{tabular}} & \multirow{2}{*}{\textbf{AUC}} & \multirow{2}{*}{\textbf{AUPRC}} & \multicolumn{2}{c}{\textbf{SPE @95\%}} & \multicolumn{2}{c}{\textbf{SPE @97\%}} & \multicolumn{2}{c}{\textbf{SPE @99\%}} \\ 
\cline{4-9}
 & & & \textbf{SEN} & \textbf{Th} & \textbf{SEN} & \textbf{Th} & \textbf{SEN} & \textbf{Th} \\ 
\hline \hline
1 & 0.852 & 0.979 & 0.436 & 0.784 & 0.342 & 0.834 & 0.191 & 0.900 \\
3 & 0.897 & 0.962 & 0.571 & 0.780 & 0.488 & 0.845 & 0.300 & 0.921 \\
5 & 0.905 & 0.946 & \cellcolor[gray]{0.9}{0.604} & \cellcolor[gray]{0.9}{0.776} & \cellcolor[gray]{0.9}{0.518} & \cellcolor[gray]{0.9}{0.852} & \cellcolor[gray]{0.9}{0.353} & \cellcolor[gray]{0.9}{0.929} \\
\hline \hline
\end{tabular}
}
\end{table}

% Old results
\iffalse
\begin{table}[htp]
\centering
\caption{Segment-level combined binary artifact detection results.}
\label{tab:segment_artifact_results_binary_all}
\scalebox{0.8}{
\centering
\begin{tabular}{ccccccccc} 
\hline \hline
\multirow{2}{*}{\begin{tabular}[c]{@{}c@{}}\textbf{Window}\\\textbf{Length}\end{tabular}} & \multirow{2}{*}{\textbf{AUC}} & \multirow{2}{*}{\textbf{AUPRC}} & \multicolumn{2}{c}{\textbf{SPE @95\%}} & \multicolumn{2}{c}{\textbf{SPE @97\%}} & \multicolumn{2}{c}{\textbf{SPE @99\%}} \\ 
\cline{4-9}
 & & & \textbf{SEN} & \textbf{Th} & \textbf{SEN} & \textbf{Th} & \textbf{SEN} & \textbf{Th} \\ 
\hline \hline
1 & 0.876 & 0.866 & 0.401 & 0.821 & 0.306 & 0.855 & 0.127 & 0.916 \\
3 & 0.870 & 0.796 & \cellcolor[gray]{0.9}{0.420} & \cellcolor[gray]{0.9}{0.812} & \cellcolor[gray]{0.9}{0.320} & \cellcolor[gray]{0.9}{0.871} & 0.114 & 0.927 \\
5 & 0.873 & 0.732 & 0.367 & 0.827 & 0.269 & 0.853 & \cellcolor[gray]{0.9}{0.133} & \cellcolor[gray]{0.9}{0.899} \\
\hline \hline
\end{tabular}
}
\end{table}
\fi

\section{Discussion}
In the following, we compare our results to the literature. Roy performed multi-class artifact classification and reported a SEN of 72.39\% for the background class~\cite{roy2019machine}. Abdi \textit{et al.} deployed wavelet transform to detect and reject artifacts. They measured the effectiveness of their artifact detector indirectly by performing brain-computer interface (BCI) classification and achieved an ACC improvement from 63\% to 72.5\% with the artifact rejection module~\cite{abdi2021eeg}. Mashhadi \textit{et al.} reject ocular artifacts by means of U-NET, and reported a mean square error (MSE) of 0.00712~\cite{mashhadi2020deep}. Meanwhile, Dhindsa performed artifact classification on an EEG dataset with four channels and achieved an ACC of 93.3\% and AUC of 0.923~\cite{dhindsa2017filter}. Finally, Pion-Tonachini \textit{et al.} deployed ICA and CNN/GAN to classify EEG independent components (IC), and achieved BAC of 0.855, 0.623, and 0.597, for 2, 5, and 7-class classification, respectively~\cite{pion2019iclabel}. For all scenarios, they reported SEN of 73\% for the background class. 

%The different studies for artifact detection reported in the literature considered various datasets and assessment metrics, thus making comparison challenging. 

Compared to these studies, our system performs better in terms of SPE, as our system reports a high SPE of 95\% while achieving decent SEN of 57.1\% for the artifact class, making it suitable for real-world application. In contrast, the studies by~\cite{roy2019machine, abdi2021eeg, mashhadi2020deep, pion2019iclabel} might be less suitable for real-world applications which require high SPE to avoid rejecting too much clean EEG. The majority of existing studies reported low SPE for the clean EEG (less than 75\%), which is unacceptable as it can lead to a significant loss of valuable EEG information.

%None of the existing works (except {Roy~\cite{roy2019machine}}) considered the TUH-ART dataset. 

%In another line of research, Abdi \textit{et al.} and Mashhadi \textit{et al.} deployed a semi-simulated dataset by injecting noise into regular EEGs to generate an artifact dataset~\cite{abdi2021eeg, mashhadi2020deep}.

%However, we noted that they trained a separate model for every channel (19 channels), even when they only deployed a small dataset (54 30s segments). They did not report the original signal mean square amplitude (MSA), so it is difficult to perceive the actual effectiveness. Finally, for the background signals, the predicted and target signals are dissimilar even with low MSE; their approach alters the EEG morphology, even if they do not contain artifacts.

\section{Conclusion}
We have proposed a neural system for automated detection of five artifact classes: chewing, electrode pop, eye movement, muscle, and shiver artifacts. The channel-wise detector consists of a CNN followed by a transformer optimized via a BM loss. The outputs of the CNN-TRF at multiple channels are then combined via another classifier for artifact detection in multi-channel EEG segments. We evaluated the segment-level detector via binary, multi-class, and multi-class multi-label classification. Finally, we implemented a non-artifact versus artifact combined binary classification for a general artifact rejection system. The proposed system can reject a substantial fraction of artifacts while only removing a small fraction of clean EEG, thus potentially improving the readability of EEG recordings. In future work, the artifact rejection can be implemented with other EEG abnormality detectors, such as seizure or slowing detectors, to enhance the performance of those abnormalities detectors.

\addtolength{\textheight}{-12cm} 

%%%%%%%%%%%%%%%%%%%%%%%%%%%%%%%%%%%%%%%%%%%%%%%%%%%%%%%%%%%%%%%%%%%%%%%%%%%%%%%%
%\section*{APPENDIX}


%\section{ACKNOWLEDGMENT}

%%%%%%%%%%%%%%%%%%%%%%%%%%%%%%%%%%%%%%%%%%%%%%%%%%%%%%%%%%%%%%%%%%%%%%%%%%%%%%%%
\bibliographystyle{elsarticle-num} 
\bibliography{refs}

\end{document}